\documentclass[twocolumn]{aastex631}
\usepackage{amsmath}
\usepackage{amssymb}
\usepackage{hyperref}
\usepackage{soul}
\usepackage{graphicx} 
\usepackage{subfigure}

\begin{document}

\title{On the origin of  $\sim 100 {\rm TeV}$ neutrinos from the Seyfert galaxy NGC 7469}

\author[0009-0003-7748-3733]{Qi-Rui Yang}
\affiliation{School of Astronomy and Space Science, Nanjing University, Nanjing 210023, China}
\affiliation{Key laboratory of Modern Astronomy and Astrophysics (Nanjing University), Ministry of Education, Nanjing 210023, China}

\author[0000-0003-4907-6666]{Xiao-Bin Chen}
\affiliation{School of Astronomy and Space Science, Nanjing University, Nanjing 210023, China}
\affiliation{Key laboratory of Modern Astronomy and Astrophysics (Nanjing University), Ministry of Education, Nanjing 210023, China}

\author[0000-0003-1576-0961]{Ruo-Yu Liu}
\affiliation{School of Astronomy and Space Science, Nanjing University, Nanjing 210023, China}
\affiliation{Key laboratory of Modern Astronomy and Astrophysics (Nanjing University), Ministry of Education, Nanjing 210023, China}
\affiliation{Tianfu Cosmic Ray Research Center, Chengdu 610000, Sichuan, China}

\author[0000-0002-5881-335X]{Xiang-Yu Wang}
\affiliation{School of Astronomy and Space Science, Nanjing University, Nanjing 210023, China}
\affiliation{Key laboratory of Modern Astronomy and Astrophysics (Nanjing University), Ministry of Education, Nanjing 210023, China}
\affiliation{Tianfu Cosmic Ray Research Center, Chengdu 610000, Sichuan, China}

\author[0000-0002-5881-335X]{Martin  Lemoine}
\affiliation{Astroparticule et Cosmologie (APC), CNRS – Université Paris Cité, 75013 Paris, France}

\begin{abstract}
The origin of TeV-PeV neutrinos detected by IceCube remains largely unknown. The most significant individual neutrino source is the close-by Seyfert galaxy NGC 1068 at 4.2$\sigma$ level with a soft spectral index. Another notable candidate is the Seyfert galaxy NGC 7469, which has been recently proposed as a potential neutrino
emitter.  The likelihood fit of the IceCube data for this source returned a very hard spectral index of $\sim 1.9$ and the excess is  dominated by two high-energy events,  issued as two neutrino alerts IC220424A and IC230416A. 
The energies of the two neutrinos are estimated to be $100-200\,$TeV, implying a maximum proton energy $E_{p,{\rm max}}>2\,{\rm {PeV}}$, significantly higher than that in NGC 1068. The lack of lower-energy neutrinos from NGC 7469 also suggests a neutrino spectrum harder than that of NGC 1068. In this paper, we analyze the {\it Fermi}-LAT observations of NGC 7469, which yield non-detection. 
By requiring the cascade flux accompanying neutrino production not to exceed the upper limit of the GeV flux,  the size of the neutrino-emitting region can be constrained when the neutrino flux takes a high value of the allowed range. 
We suggest that cosmic ray protons are accelerated to PeV energies via turbulence or magnetic reconnection in the corona of NGC 7469 and interact with optical and ultraviolet (OUV) photons from the accretion disk and X-rays from the corona through the $p\gamma$ process, producing neutrinos with energy of $100-200$ TeV.
In the turbulence acceleration scenario, the required maximum proton energy can be achieved with a magnetization parameter close to unity ($\sigma\sim 1$), while in the reconnection scenario, a magnetization parameter with $\sigma\sim 10$ is needed. In both scenarios, a pair dominated composition for the corona is preferred. 
The difference in the neutrino spectrum between NGC 7469 and NGC 1068 could be due to a different magnetization parameter despite the fact that they belong to the same type of AGN. 
\end{abstract}

\section{Introduction} 

The IceCube collaboration \citep{2022Sci...378..538I} reported an excess of neutrino
events associated with NGC 1068, a nearby type-2 Seyfert galaxy, with a significance of 4.2$\sigma$. The
reported neutrino flux is significantly higher than the GeV gamma-ray flux of the galaxy \citep{Ackermann2012ApJ...750....3A}.  
It has hence been suggested that the opaque cores of active galactic nucleus (AGN)
can be the high-energy neutrino sources, where
dense radiation attenuates gamma-rays while providing abundant targets for neutrino production \citep{Murase2020ApJ...902..108M,Murase2022ApJ...941L..17M,Inoue2020ApJ...891L..33I,Kheirandish2021ApJ...922...45K,Halzen2022arXiv220200694H,Kurahashi2022ARNPS..72..365K,Eichmann2022ApJ...939...43E,Halzen2023arXiv230507086H}. Specifically, magnetized corona
are suggested to be promising proton accelerators that can  produce neutrinos mainly through interactions with coronal X-rays \citep{Stecker1991PhRvL..66.2697S,Inoue2019ApJ...880...40I,Murase2022ApJ...941L..17M}. 
Neutrino emission from AGN cores directly points to the
presence of a relativistic proton population, but the specific
particle acceleration mechanism at work remains an open
question. Stochastic turbulence acceleration and magnetic
reconnection may naturally coexist in magnetized plasmas. 
The two processes can lead to different proton spectrum and the maximum proton energy, so the neutrino emission could, in principle, provide a probe of the particle acceleration mechanism and the underlying physical condition.  

The neutrino emission of NGC 1068 {suggests} a soft spectrum $d \phi_{\nu}/d E_{\nu} \propto  E_{\nu}^{-3.4\pm0.2}$ in 1.5–15 TeV. {Taken at face value, this spectrum suggests that the parent proton spectrum is intrinsically soft in the corresponding region of proton energies around $30-300\,$TeV. However, tight constraints on the overall energy budget then require a break at some energy, with a harder spectrum below the break~\citep{Kheirandish2021ApJ...922...45K}. Various models have been considered to reproduce the requisite broken power-law spectral shape. For instance, the soft spectrum at high energies might correspond to the tail at and around the maximal energy in the accelerator~\citep{Murase2020PhRvL.125a1101M,Inoue2020ApJ...891L..33I,Eichmann2022ApJ...939...43E,Inoue2022arXiv220702097I,2025A&A...697A.124L}. Alternatively, the break could represent the maximal energy reached in a reconnection layer, before re-acceleration in the surrounding turbulence~\citep{Mbarek2024PhRvD.109j1306M}. The soft high-energy spectral part could also result from acceleration in inhomogeneous turbulence, or re-acceleration in a shear layer at the base of the jet~\citep{2025A&A...697A.124L}, or from the effect of a guide field in reconnection scenarios~\citep{Fiorillo2024ApJ...961L..14F}.}

NGC 7469 is a  Seyfert 1.2 galaxy located at a redshift of $z \approx 0.016$, hosting a supermassive black hole with mass $M_{\rm BH} = 9\times 10^6 M_{\odot}$ \citep{BK2015PASP..127...67B}. The hydrogen column density for NGC 7469 is approximately $N_H \sim 10^{20}\, \rm{cm^{-2}}$, which suggests it is oriented face-on and unobscured. NGC 7469 has recently been proposed as a potential neutrino emitter following the detection of two track-like neutrino events (IC220424A and IC230416A) with energy of $\sim 184\, {\rm TeV}$ and  $\sim 127\, {\rm TeV}$, respectively, within its vicinity \citep{Sommani2025ApJ...981..103S}. 
A recent IceCube analysis \citep{IceCube2025arXiv251013403A} for neutrino emission from a list of 47 X-ray bright, non-blazar AGNs found NGC 7469 as the
most significant source, excluding NGC 1068, with a local significance of $3.8\sigma$ and a global significance of  $2.4\sigma$. 
The likelihood fit for this source returned a hard spectral index of $\sim 1.9$ and the excess is fully dominated by two high-energy events, IC220424A and IC230416A.

The neutrino energy, which is significantly higher than that of NGC 1068, implies that the maximum proton energy should be  $E_{p,{\rm max}}\ge 2\, {\rm PeV}$.  The non-detection of TeV neutrinos from NGC 7469, despite the presence of two 100 TeV events, suggests a hard neutrino spectrum ($s_\nu \lesssim 2$). 
In this paper,  we aim to study the origin of the high-energy neutrinos from  NGC 7469, focusing on explaining this hard neutrino spectrum of NGC 7469.

\section{GeV flux limit and the neutrino emission of NGC 7469}
\subsection{{\it Fermi}-LAT observations of NGC 7469}
We used 17 years of {\it Fermi}-LAT data from August 2008 to August 2025 to study the GeV emission of NGC 7469. We select an energy range of 100 MeV–1 TeV, and bin the data using eight energy bins per decade. We select photons corresponding to the {\tt\string P8R3\_SOURCE\_V3} instrument response and event type FRONT + BACK (evtpye = 3) are used. 
To account for the diffuse emission, we modeled the
Galactic diffuse emission model ({\tt\string gll\_iem\_v07.fits}) with isotropic component({\tt\string iso\_P8R3\_SOURCE\_V3\_v1.txt}) relevant to the {$\tt\string SOURCE$} event class. We use recommended time selection of $\rm (DATA\_QUAL>0)\&\&(LAT\_CONFIG == 1)$.
To minimize the contamination from the Earth limb, the maximum zenith angle is set to be $90^{\circ}$. 
The analysis is performed using the publicly available software $\textit{Fermitools}$ (v2.2.0) and the $\textit{Fermipy}$ tool (version 1.2.2) \citep{Fermipy2017Wood}.

The data selection was within a region of interest (ROI) of $20^\circ$ around NGC 7469 at (R.A., Dec.) = (345.82$^\circ$, 8.87$^\circ$). We include the Galactic diffuse emission (GDE), isotropic emission and all sources listed in the fourth {\it Fermi}-LAT catalog \citep{Fermi4FGL-DR4_2023} in the background model. All sources within $8^{\circ}$ of the center, as well as the GDE and isotropic emission components, are left free. Conducting a binned maximum-likelihood analysis, NGC 7469 is not detected (TS = 0). So we calculate the 95\% confidence-level upper limits on the gamma-ray flux from NGC 7469, which are summarized in Table \ref{tab:fermiNGC7469}.

\begin{table}[ht]
\centering
\caption{{\it Fermi}-LAT flux  upper limit of NGC 7469.} \label{tab:fermiNGC7469}
\begin{tabular}{cccc}
\hline\hline
Energy Low & Energy High & 95\% Flux UL  \\
{[GeV]} & {[GeV]} & ($\times10^{-8}$ MeV cm$^{-2}$ s$^{-1}$)  \\
\hline
 0.1   & 0.3   &   6.02   \\
 0.3   & 1.0   &  5.72   \\
 1.0   & 3.16  &  5.18   \\
 3.16  & 10    &  7.49   \\
 10    & 31.6  &  3.45   \\
 31.6  & 100   &  12.0   \\
 100   & 1000  &  29.2   \\
\hline
\end{tabular}
\end{table}

\subsection{IceCube observations of NGC 7469}
The recent IceCube analysis of NGC 7469 tested two spectral hypotheses. The likelihood fit for this
source returned a hard spectral index of $s_{\nu}\sim 1.9$ and the excess is fully dominated by two high-energy events, IC220424A and IC230416A. \citep{IceCube2025arXiv251013403A}
The event IC220424A was classified as a gold alert with a signalness of 50\% and an energy of 184 TeV. IC230416A was classified as a bronze alert with a signalness of 34\% and an energy of 127 TeV. If either one or both neutrinos originated from the source, the maximum proton energy should be at least 
\begin{equation}
    E_{p,{\rm max}}\gtrsim 20\, E_\nu\simeq 2\, {\rm PeV}.
\end{equation}
As reported by  IceCube\citep{IceCube2025arXiv251013403A}, the all-flavor neutrino flux has a large uncertainty range, from $E_{\nu}  F_{\nu}\sim10^{-14} \, {\rm erg\,cm^{-2}\,s^{-1}}$ to $E_{\nu}  F_{\nu}\sim10^{-12} \, {\rm erg\,cm^{-2}\,s^{-1}}$.  
The neutrino luminosity is approximately $L_{\nu}=4\pi D^2 (1+z)^2 E_\nu  F_\nu \simeq  10^{41}\,{\rm erg\,s^{-1}} (E_\nu  F_{\nu}/10^{-13} \, {\rm erg\,cm^{-2}\,s^{-1}}) $ for a distance of $D \approx {70}\,{\rm Mpc}$, where $z$ is the redshift.
The neutrino flux is  determined by the hadronic process efficiency $f_{p\gamma,pp}$ and cosmic ray flux as
\begin{equation}
    E_\nu F_{\nu} \approx \frac{3}{8}f_{p\gamma,pp}E_{p}F_{p}.
    \label{Eq:Neutrino spectrum}
\end{equation}
The proton luminosity $L_p$ needed to explain the $\gtrsim100$ TeV neutrino spectrum then reads
\begin{equation}
\begin{aligned}
       L_p&=4\pi D^2 (1+z)^2E_{p}{F_{p}}\\
       &=1.7\times 10^{41}\, {\rm erg\, s^{-1}} f_{p\gamma,pp}^{-1}\left(\frac{E_{\nu}  F_{\nu}}{ 10^{-13} \, {\rm erg\,cm^{-2}\,s^{-1}}}\right).
\end{aligned}
\end{equation}
The  Eddington luminosity of NGC 7469 is $1.1\times10^{45}{\rm erg s^{-1}}$ for a black hole mass of $M_{\rm BH}=9\times10^6M_\odot$. The bolometric luminosity of the accretion disk
($L_{\rm disk}$) is estimated to be $20-30\%$ of the Eddington luminosity, i.e., $L_{\rm bol}\simeq3.5\times10^{44}\,{\rm erg\, s^{-1}}$
\citep{Mehdipour2018A&A...615A..72M,Partington2025ApJ...986...81P}. {The required proton luminosity thus corresponds to  a modest fraction of the bolometric luminosity, as long as the neutrino production efficiency is not too low.}

\subsection{Neutrino–Gamma-ray connection}
High-energy neutrinos are produced through $p\gamma$ and/or $pp$ interactions. In either case, the neutrino emission must be accompanied by gamma-ray production and their fluxes are comparable. The 95\% confidence-level upper limits on gamma-ray flux of NGC 7469 at GeV-TeV range measured by {\it Fermi}-LAT is $E_{\gamma}F_{\gamma} \sim 10^{-13} \, {\rm erg\,cm^{-2}\,s^{-1}} $. If the neutrino flux approaches the upper bound of the measured flux range, i.e., $ E_{\nu} F_\nu \sim 10^{-12}\,{\rm erg\, cm^{-2}\, s^{-1}} $ around 100 TeV, the neutrino emitting region should be optically thick, where gamma-ray emission is absorbed, while neutrinos can emit.  In addition, NGC 7469 has a very low level radio emission, indicating no strong  jet-ejection event \citep{Seifina2018A&A...619A..21S}. Therefore, unlike TXS 0506+056 in which the neutrinos can arise from  the jet, high-energy cosmic rays and neutrinos of NGC 7469 are most likely produced in the optically thick corona.
The possible acceleration mechanisms include turbulence acceleration and magnetic  reconnection  acceleration.


\section{Acceleration in magnetized turbulence}
\subsection {Physical parameters}
We examine here the possibility of accelerating protons up to PeV energies, through stochastic acceleration in the turbulent corona of NGC~7469. 
We consider a generic setup in which a supermassive black hole with a mass of $M_{\rm BH}$ and gravitation radius  $R_{\rm g}$ is embedded in the luminous radiation of accretion disk and corona. For simplicity, the corona that encompasses the inner accretion disk is assumed to be quasi-spherical and compact, with a characteristic length of
$R_c\sim 10 R_{\rm g}$ . 
The total number density of electrons and positrons, ${n}_{e}$, is self-regulated such that $\tau_{\rm T}\sim 1$ \citep{Prince_2025MNRAS.tmp..960P}, resulting in 
\begin{equation}
{n_{e}}\sim \tau_{\rm T}/\sigma_T R_c \sim 10^{11}\left(\frac{10 R_{\rm g}}{R_c}\right)\, \rm{cm}^{-3}. 
\label{Eq: electron number density}
\end{equation}
We anticipate that the number of pairs generated in the corona could exceed the number of primary electrons, in which case ${n_p}<{n_e}$, where $n_p$ is the number density of thermal protons. We thus keep the ratio $n_p/n_e$ as a free parameter in the following. The characteristic proton temperature is estimated to lie close to the virial temperature, $T_p = GM_{\rm BH}m_p/3kR_c  \simeq3\times 10^{11}\,{\rm K}(10R_{\rm g}/R_c)$, where $k$ is Boltzmann constant. {Throughout, we set $M_{\rm BH}=9\times10^6 M_{\odot}$}. While still high, radiative cooling results in electron temperatures that are significantly lower ($T_e<T_p$), as evidenced by the observed X-ray spectral cutoffs at hundreds of keV
(e.g., \cite{Kamraj2018ApJ...866..124K,Kamraj2022ApJ...927...42K,Kammoun2024FrASS..1008056K}). In the case of NGC~7469, observations indicate that the corona generates an X-ray luminosity of $L_{2-10\,\rm{keV}} = 1.5\times 10^{43}\,\rm{erg\,s^{-1}}$, with X-ray photon index of $s_X = 1.9$ and cut-off energy of 284 keV \citep{Prince_2025MNRAS.tmp..960P}.

To estimate the magnetic field strength, we relate the magnetic field strength to the plasma pressure through the $\beta_p$ parameter (defined as the ratio between the thermal plasma pressure and the magnetic pressure),  expected to be $\sim 0.1-1$ (for recent numerical simulations, see in particular \citet{Liska2022ApJ...935L...1L}.  This gives $B=\sqrt{8\pi n_pkT_p/\beta_p}\simeq  1.2\times 10^4\,{\rm G}(n_p/n_e)^{1/2}\beta_p^{-1/2}(10R_{\rm g}/R_c)^{1/2}$, assuming that the proton pressure dominates over the leptonic component. 
Correspondingly, the turbulent Alfvénic velocity $v_{\rm A}$ that controls the acceleration timescale, is given by $v_{\rm A} = \sqrt{2} \beta_p^{-1/2} (k_B T_p/m_p c^2)^{1/2} \simeq 0.3\,c\,\beta_p^{-1/2}(10 R_{\rm g}/R_c)^{1/2}$.

We obtain similar results, albeit with different parameter scalings, if we
assume that the X-ray luminosity results from instantaneous dissipation of the turbulent magnetic energy density, following~\citet{Beloborodov2017ApJ...850..141B}, \citet{2024PhRvL.132h5202G} and \citet{Fiorillo2024ApJ...974...75F}. This implies $L_X \simeq u_B V_c \times 0.1v_{\rm A}/\ell_c$, where $V_c$ denotes the coronal volume, $v_{\rm A}$ the Alfvénic velocity, $\ell_c$ the outer scale of the turbulence, and $\ell_c/(0.1v_{\rm A})$ representing the mean dissipation time, also corresponding to a few eddy turn-over times. Using $v_{\rm A} = B/(4\pi n_pm_p c^2)^{1/2}$, assuming $n_p m_p > n_e m_e$ and inverting the above, leads to $B \,\simeq\, 9\times 10^3\,{\rm G}$ for $L_X=1.5\times 10^{43}\,$erg/s, $n_p/n_e\sim 1$, $\ell_c\sim R_{\rm g}$ and $R_c \sim 10R_{\rm g}$. The corresponding Alfv\'enic velocity reads $v_{\rm A}\,\simeq\,0.2\,c$ for these same values.

Particle acceleration in magnetically-dominated turbulence is mainly controlled by the amplitude of the turbulence on the outer scale $\ell_c$, e.g., the relative magnetic perturbation $\delta B/B$, with $B$ the mean total magnetic field, and the characteristic eddy velocity on that scale and the turbulent Alfvénic velocity $v_{\rm A}$. We assume here $\delta B/B \sim 1$, noting that otherwise, the acceleration timescale would likely become prohibitive with respect to proton acceleration up to $\sim O({\rm PeV})$ energies (see also thereafter). This regime of large-amplitude, semi-relativistic turbulence has recently gained insight through large-scale particle-in-cell (PIC) numerical experiments~\citep[e.g.][]{Zhdankin2018ApJ...867L..18Z,ComissoSironi2019ApJ...886..122C,Wong2020ApJ...893L...7W,Bresci+22,2023ApJ...944..122M,2025MNRAS.543.1842W,2025arXiv250604212D}, and analytical developments~\citep[e.g.][]{2021PhRvD.104f3020L,2022PhRvL.129u5101L}. The mean energy diffusion coefficient measured in the above simulations is $D_{\gamma\gamma}\approx 0.1\, \gamma^2\,\sigma c/\ell_{\rm c}$, where $\sigma$ is the magnetization parameter, defined as the ratio of magnetic to plasma energy densities as $\sigma = B^2/( 4\pi n_p m_p c^2+4\pi n_e m_e c^2)\simeq B^2/( 4\pi n_p m_p c^2)$ since $n_pm_p>n_e m_e $. In terms of $v_{\rm A}$, $\sigma = \beta_{\rm A}^2/(1-\beta_{\rm A}^2)$, with $\beta_A\equiv v_{\rm A}/c$. 

\subsection {Maximum proton energy}
The turbulence acceleration timescale is given by 
\begin{equation}
    t_{\rm tur } = \frac{1}{4}\frac{\gamma^2}{D_{\gamma\gamma}} \simeq 2.5\,\sigma^{-1}\frac{\ell_c}{c} \approx 110 \,{\rm s}\,\sigma^{-1}\left(\frac{\ell_c}{R_{\rm g}}\right).
\label{Eq: Turbulence Acc}
\end{equation}

The cooling processes, including $pp$, $p\gamma$, Bethe-Heitler, proton synchrotron are taken into account(see details in Appendix.\ref{Sec:Appendix A}). Among them,  $p\gamma$ dominates the cooling at PeV energy band, as shown in the section \ref{section: Neutrino spectrum}. The timescale of $p\gamma$ cooling is estimated by
\begin{equation}
\begin{aligned}
    t_{p\gamma}\approx \left(n_{\gamma}\hat{\sigma}_{p\gamma}\kappa_{p\gamma}c\right)^{-1},
    \label{Eq:timescale pgamma}
\end{aligned}    
\end{equation}
where $\hat{\sigma}_{p\gamma} \approx 5\times 10^{-28}~\rm{cm}^{2}$ is the  cross section for the photomeson process, $\kappa_{p\gamma} \sim 0.2$ is the inelasticity for $p\gamma$, $n_{\gamma} = L_{\varepsilon_{\gamma}}/4\pi R^2c\varepsilon_{\gamma} $ is the number density of the target photon field with the luminosity  $L_{\varepsilon_\gamma}$  at photon energy $\varepsilon_\gamma$ and $R$ is the radius of emitting region. The energy of target photon $\varepsilon_\gamma$  is related to the proton energy via the $\Delta$-resonance threshold condition $E_p \varepsilon_\gamma \sim 0.15\,{\rm GeV}^2$. 
By equating the timescales of turbulence acceleration Eq.\eqref{Eq: Turbulence Acc} and $p\gamma$ cooling Eq.\eqref{Eq:timescale pgamma}, we can deduce the maximum energy for protons accelerated in the coronal turbulence,  
\begin{align}
    E_{p,{\rm tur, max}} &\,=\,\frac{0.6\pi R_c^2c \sigma\, {\rm GeV}^2}{5\hat{\sigma}_{p\gamma}\kappa_{p\gamma}\ell_c L_{\varepsilon_{\gamma}}}\nonumber\\
    &\simeq 2\, {\rm PeV}\,\sigma\left(\frac{R_c}{10R_{\rm g}}\right)^2\left(\frac{R_{\rm g}}{\ell_c}\right)\left(\frac{10^{41} \,{\rm erg\,s^{-1}}}{L_{\varepsilon_{\gamma}} }\right),
    \label{Eq:Tur Epmax}
\end{align}
where $\varepsilon_{\gamma}\simeq 0.15 \,{\rm GeV}^2 / E_{p,{\rm tur, max}} \simeq 75\,{\rm eV}$ denotes the target photon energy corresponding to $E_{p,{\rm tur, max}} \simeq 2\,{\rm PeV}$. The luminosity at this photon energy is $L_{\varepsilon_{\gamma}} \simeq 10^{41} \,{\rm erg\,s^{-1}}$, which arises from the “Big Blue Bump” (BBB) and X-ray component, as characterized by \cite{Prince_2025MNRAS.tmp..960P}. 

To account for neutrinos with energy $E_{\nu}\gtrsim 100 \, {\rm TeV}$,  a relatively high magnetization of $\sigma \sim 1 $ is necessary, i.e. a relativistic Alfv{\'e}n speed $v_A\sim c$, corresponding to $\beta_p \simeq 0.1$. In principle, the magnetization $\sigma$ that controls $E_{p,{\rm tur, max}}$ should be multiplied by a factor of ($\delta B/B)^2$ to represent the effective turbulent magnetization in magnetically dominated turbulence. If $\delta B/B < 1$, the acceleration would be suppressed, making it difficult for protons to reach PeV energies. Therefore, $\delta B/B \sim 1$ is necessary in our scenario.

\begin{figure}
\centering
\subfigure{
\includegraphics[width=0.45\textwidth]{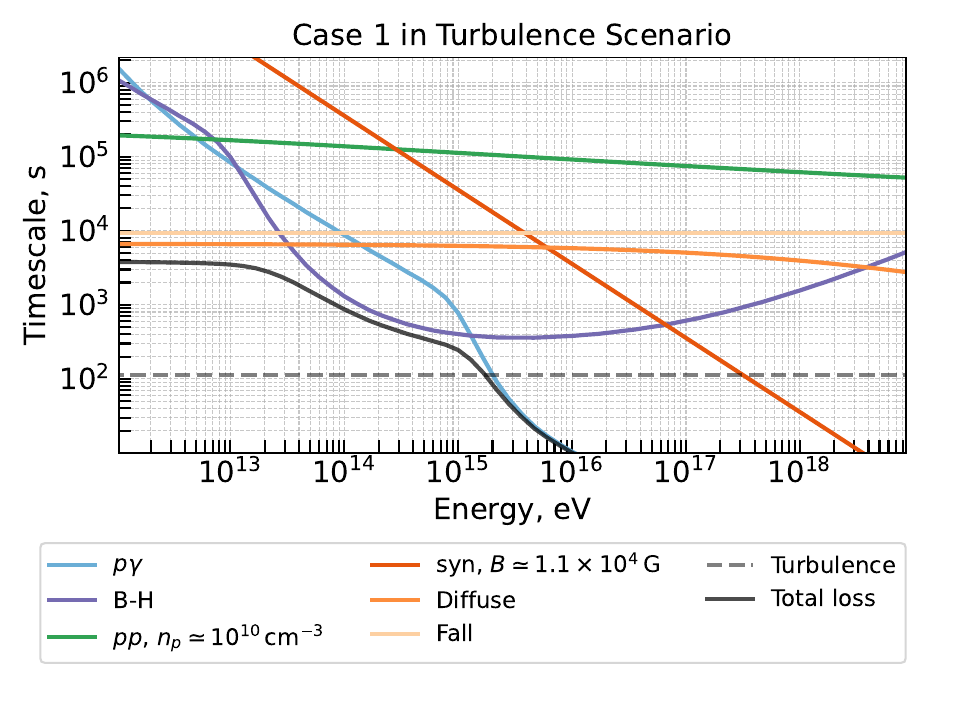}}
\subfigure{
\includegraphics[width=0.45\textwidth]{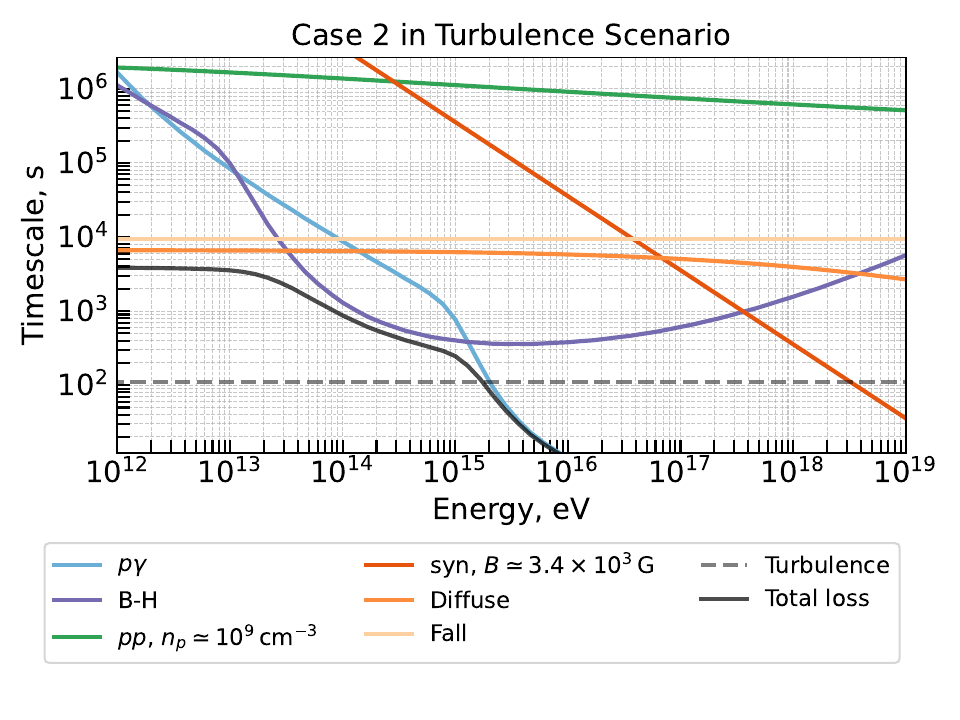}}

\caption{
Timescales of proton acceleration, escape, and cooling in the corona of NGC 7469 in turbulence scenario. The upper panel and lower panel represent the case 1 and case 2 in turbulence scenario, respectively. The timescale of turbulence acceleration is shown in black dashed line and the thin solid colored line refers to the energy-loss timescale. 
The shared parameters for the two cases are $R_c \approx 10 R_{\rm g}$, $\beta_p\simeq 0.1$, $\ell_c = R_{\rm g}$. In case 1, $n_{p} \simeq10^{10}\,{\rm cm^{-3}}$, and the magnetic field is  $B = 1.1\times 10^4\,{\rm G}$; In case 2, $n_{p} \simeq 10^{9}\,{\rm cm^{-3}}$, and the magnetic field is  $B = 3.4\times 10^3\,{\rm G}$.
 }
\label{Fig:Timescale_Turbulence}
\end{figure}

\subsection {Spectral shape}

To estimate the spectral shape, we follow here the recent discussion of \citet{2025A&A...697A.124L}. If particles are injected continuously then accelerated in the turbulent corona, the low-energy part of the proton spectrum scales as ${\rm d}n_p/{\rm d}\gamma\propto \gamma^{-s}$ with $s\gtrsim 1$. Specifically, $s=1 + t_{\rm tur}/t_{\rm esc}$ for $D_{\gamma\gamma}\propto \gamma^2$ if the escape timescale $t_{\rm esc}$ is independent of energy, as $t_{\rm tur}$. This is notably the case at low energies, as particles are then transported by turbulent diffusion. Nevertheless, in the present case $t_{\rm tur}/t_{\rm esc}$ is small compared to unity, so that this correction is a small effect. At high energies, the spectrum will either cut-off because of a finite acceleration timescale, set by the infall time into the BH (or advection time out of the corona by an outflow), because of energy losses, or it will transit to $s\approx 2$ because of nonlinear feedback of the accelerated particles on the turbulence. This effect arises once the energy rate at which high-energy protons draw energy from the cascade becomes commensurable with the rate at which energy is injected into the turbulent cascade. In practice, this feedback binds the high-energy proton energy density $u_p$ to values comparable to the turbulent energy density $u_B$ for $v_{\rm A}/c\sim O(1)$. 

The proton energy density is derived from the neutrino flux as
\begin{align}
    u_p&\,\simeq\,\frac{8E_{\nu}L_{\nu}t_{p\rightarrow\nu}}{3V_c}C_p,\nonumber\\
    &\,\simeq\, 1.6\times10^4 \,{\rm erg\,cm^{-3}}\,\,\left(\frac{E_{\nu}F_{\nu}}{ 10^{-13}\,{\rm erg\,cm^{-2}\,s^{-1}}}\right)\nonumber\\
    &\quad\quad\times\left(\frac{t_{p\rightarrow\nu}}{10^3\,{\rm s}}\right)\left(\frac{10 R_{\rm g}}{R_c}\right)^3C_p\,,
    \label{eq:up}
\end{align}

where $E_{\nu}\simeq 100\,{\rm TeV}$, $F_{\nu}$ denotes the differential neutrino flux at that energy, $t_{p\rightarrow\nu}$ denotes the hadronic loss timescale to neutrino production, which corresponds to the harmonic mean of $t_{p\gamma}$ and $t_{pp}$, and $V_c$ denotes the coronal volume. $C_p$ is the correction factor from the differential luminosity to the integrated proton luminosity. 
By comparison, the turbulent magnetic energy density reads 
{$u_B \sim 10^6\,{\rm erg/cm^3}$ for $B=5\times 10^3\,$G}, 
which is much larger than the proton energy density even for the highest allowed neutrino flux of NGC 7469 for the quoted fiducial values of the  parameters, in particular for $n_p/n_e \sim 1$. 

In the case of $u_p \ll u_B$, the turbulence would  generate a hard proton spectrum with spectral index of $s\simeq 1$, extending up to the energy where losses (cooling or escape) cut off the spectrum. However, at an energy $\sim 1\,$PeV, the above energy density corresponds to a number density $\sim 10\,$cm$^{-3}$. If $s=1$, the number density of non-thermal protons is $Edn/dE\propto E^0$, so the total number density of non-thermal protons from low to high energies would be $\ O (100)\,$cm$^{-3}$. This would imply that the fraction of particles injected out of the thermal pool into the acceleration process is a tiny number compared to unity, unlike what is seen in PIC simulations at $\sigma\sim 1$.  Therefore we discard in the following a spectral shape characterized by $s\simeq 1$ from injection $\sim\ O({\rm GeV})$ to the maximum energies.

If so, the proton spectrum must be significantly softer than $\simeq 1$, such as produced by particle feedback on the turbulence. The non-thermal proton energy density must then be comparable to the magnetic energy density, namely $u_p\sim u_B$. This condition can be satisfied if the proton number density in the coronal thermal pool is smaller than that of electrons (or pairs). In effect, setting $u_p\sim u_B$, the thermal proton number density is estimated to be
\begin{align}
   n_p&\simeq \frac{\beta_p u_B}{kT_p}\sim \frac{8\beta_pE_{\nu}L_{\nu}t_{p\rightarrow\nu}C_p}{3V_ck T_p} \nonumber\\
    &= 5\times10^{8}\,{\rm cm^{-3}}\,\left(\frac{\beta_p}{0.1}\right) \left(\frac{E_{\nu}F_{\nu}}{10^{-13}\,{\rm erg\,cm^{-2}\,s^{-1}}}\right) \left(\frac{t_{p\rightarrow\nu}}{10^3\,{\rm s}}\right) \nonumber\\
    &\quad\quad\times\left(\frac{10 R_{\rm g}}{R_c}\right)^2\left(\frac{C_p}{10}\right).
    \label{Eq:np_tur}
\end{align}
This number density is smaller than the number density of electrons given by Eq.\ref{Eq: electron number density} even if the neutrino flux takes the highest value allowed by the IceCube observations. This indicates that the composition of the corona is dominated by pairs. 

{A caveat here is that we consider a one-zone approximation. Relaxing this assumption opens up the possibility that proton acceleration up to the highest energies occurs in a region of small filling fraction, while acceleration is less efficient elsewhere, for instance because $\sigma$ takes different values in different parts of the corona~\citep{2025A&A...697A.124L}. }

{To describe the proton spectrum,} we adopt a broken power-law spectral shape, with $s\sim 1$ at low energies, $s\sim 2$ at high energies and an overall normalization $u_p \sim u_B$. The break energy can be written~\citep{Lemoine2024PhRvD.109f3006L}
\begin{equation}
\begin{aligned}
 E_{p,{\rm br}}&\simeq E_0\left(\frac{v_A}{c}\right)^{-1}\beta_p^{-1} x_{\rm nth}^{-1}\\ 
 &\simeq 10\, {\rm GeV} \left(\frac{E_0}{m_pc^2}\right)\left(\frac{v_A}{c}\right)^{-1}\left(\frac{1}{\beta_p}\right)\left(\frac{0.1}{x_{\rm nth}}\right),
\end{aligned}
\end{equation}
where $E_0 \sim m_p c^2$ denotes the energy at injection in microscopic reconnecting current sheets (with $\sigma \sim 1$) and $ x_{\rm nth}$ is the fraction of the thermal plasma  converted into non-thermal particles.

\subsection {Neutrino spectrum}
\label{section: Neutrino spectrum}
Due to the large uncertainty in the neutrino flux measured by IceCube, for the sake of simplicity,  we consider two representative cases for the neutrino flux, namely, case 1 with a high neutrino flux  of $10^{-12}\, {\rm erg\,cm^{-2}\,s^{-1}}$ and case 2 with a low neutrino flux  of $10^{-13}\, {\rm erg\,cm^{-2}\,s^{-1}}$.

We considered proton cooling processes, including $p\gamma$ process, Bethe–Heitler pair production, $pp$ collisions, and proton synchrotron, as well as escape via free-fall and diffusion, together with acceleration by turbulence. The timescales for case 1 and case 2 are shown in the upper and lower panels of Fig.~\ref{Fig:Timescale_Turbulence}, respectively.
The details of timescale calculation are presented in Appendix.\ref{Sec:Appendix A}.  In addition, the effect of pion cooling is also taken into account.
Thus,  the neutrino spectrum is calculated by 
\begin{equation}
    E_{\nu}^2\frac{dN_{\nu}}{dE_{\nu}}\simeq\frac{3}{8}f_{p\gamma}\xi_{\pi}E_{p}^2\frac{dN_p}{dE_p}\vert_{E_p\simeq20 E_{\nu}}, 
\end{equation}
where $f_{p\gamma} \simeq t_{p\gamma}^{-1}/ t_{\rm loss}^{-1}$ is the efficiency of $p\gamma$ process and $\xi_{\pi}$ is the pion cooling efficiency. $t_{\rm loss}$ is the timescale of total energy loss shown by Eq.\eqref{Eq:loss}. To account for the $\gtrsim 100 \,{\rm TeV}$ neutrino, we adopt $\sigma \sim 1$ (corresponding to $\beta_p \sim  0.1$).
The break energy is $E_{p,{\rm br} }\simeq 100\,{\rm GeV}$, and the maximum energy is adopted by $2 \,{\rm PeV}$ .

\begin{figure}
    \centering
    \subfigure{
    \includegraphics[width=0.45\textwidth]{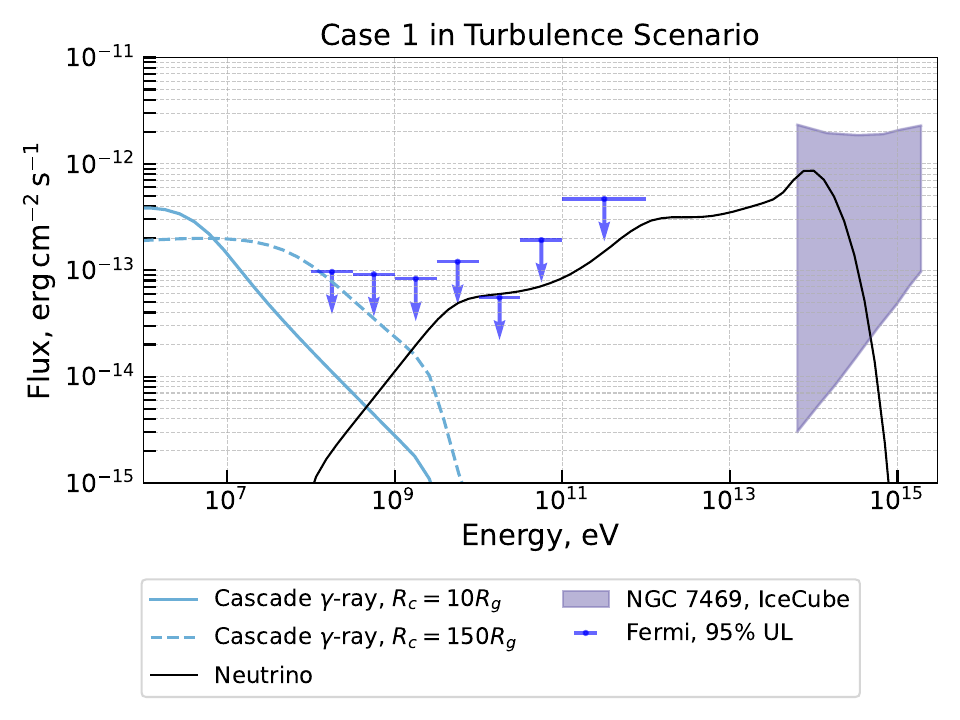}}
    \subfigure{
    \includegraphics[width=0.45\textwidth]{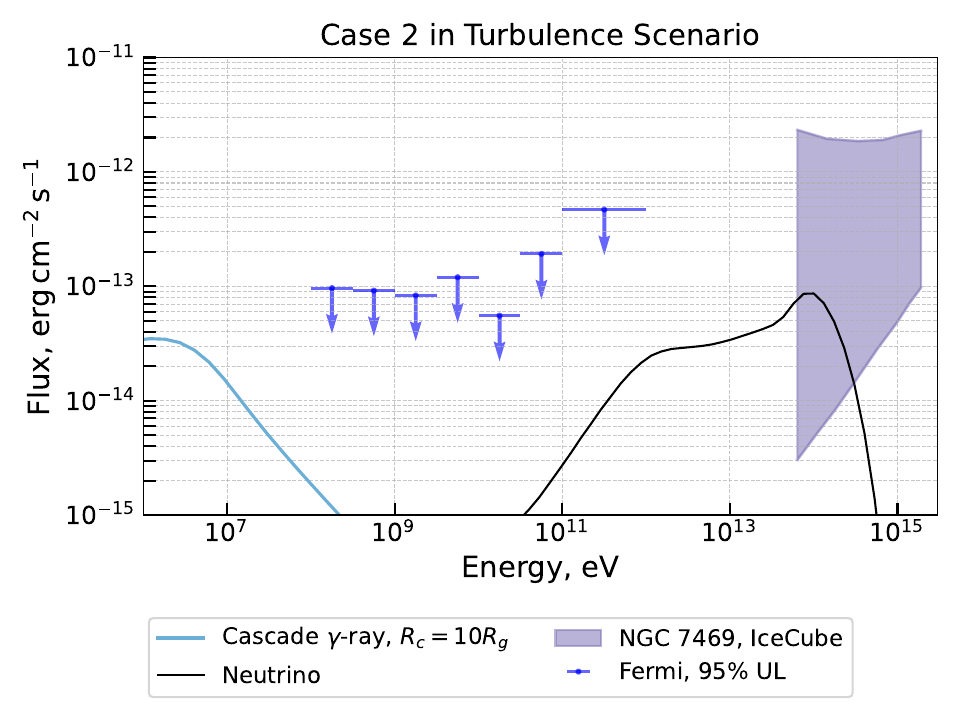}}
\caption{Multi-messenger emissions from NGC 7469 in the turbulence acceleration scenario. The upper panel illustrates the case 1 corresponding to a high neutrino flux of $10^{-12}\, {\rm erg \,cm^{-2}\,s^{-1}}$, while the lower panel shows the case 2 corresponding to a lower neutrino flux of $10^{-13} \, {\rm erg \,cm^{-2}\,s^{-1}}$. For both cases, the purple shaded region indicates the neutrino flux reported by IceCube \citep{IceCube2025arXiv251013403A}. The blue arrows indicate the upper limits for GeV gamma-ray emission imposed by {\it Fermi}-LAT.   The black solid curve represents the neutrino spectrum whereas the blue curves show the cascade gamma-ray emission,  We normalize the proton spectrum by Eq.\eqref{eq:up}. The parameters are adopted as $\beta_p \simeq 0.1$ ($\sigma \simeq 1$) with the Alfvén speed of $v_A\simeq c$ and $ E_{p,{\rm max}}\simeq2 \,{\rm PeV}$.}
    \label{Fig:Neutrino spectrum turbulence}
\end{figure}

We show the modeling of the neutrino emission of NGC 7469 and associated cascade gamma-ray emission in Fig.~\ref{Fig:Neutrino spectrum turbulence}. The upper  and lower panels represent case 1 and case 2, respectively.
The neutrino spectrum  is shown in the solid black curve and the corresponding cascade gamma-ray spectra are shown in blue curves (both dashed and solid curves). As can be seen, the neutrino spectrum appears quite hard up to the  maximum energy.
In case 1, as shown in the upper panel, the upper-limit GeV flux imposed by {\it Fermi}-LAT can constrain the size of the neutrino-emitting region to be $R_c < 150R_{\rm g}$ ($R_{\rm g}$ is the gravitational radius), by ensuring that the cascade emission (shown by the blue dashed line)  associated with neutrino emission does not exceed the observed upper limit. 
On the other hand, for case 2, the neutrino flux is comparable to the level of the GeV upper limit, so  no constraint on the size of the neutrino-emitting region can be obtained. 
The neutrino flux can also be used to  infer the  composition of the corona.  In case 1, for a neutrino flux  of $E_{\nu}F_{\nu}\sim 10^{-12}\, {\rm erg\,cm^{-2}\, s^{-1}}$, the required proton power is $L_p \simeq 5\times 10^{43}\,{\rm erg \,s^{-1}}$. The condition that an efficient feedback occurs (i.e., $u_p=u_B$) yields a coronal proton number density of $n_p\sim  10^{10}\,{\rm cm^{-3}}$ for typical parameter values of $\beta_p=0.1$ and $R_c=10R_{\rm g}$, as estimated from Eq.\ref{Eq:np_tur}.
For the case 2,  the lower neutrino flux of $E_{\nu}F_{\nu}\sim 10^{-13}\, {\rm erg\,cm^{-2}\, s^{-1}}$ results in a required proton power of $L_p \simeq  5\times 10^{42}\,{\rm erg \,s^{-1}}$.
The coronal proton number density  is $n_p\sim 10^{9}\,{\rm cm^{-3}}$ in this case.
In both cases, the proton number density is lower than that of electrons ($n_p/n_e < 1$), indicating that the corona is predominately composed of electron–positron pairs.

\section{Magnetic reconnection acceleration} \label{Sec:Magnetic reconnection acceleration}
\subsection{The magnetization parameter and the composition of the corona}

In this section, we consider the magnetic reconnection in the corona as the primary acceleration process. Following earlier studies \citep{Beloborodov2017ApJ...850..141B,SB2020ApJ...899...52S}, we assume that the X-ray energy density is a fraction of the corona magnetic energy density, shown by
\begin{equation}
    U_X \simeq \beta_{\rm rec} U_B, 
   \label{Eq: L_X}
\end{equation}
where  $U_X = L_X/(4\pi R_c^2 c)$, $U_B = B^2/8\pi$, $R_c$ is the coronal radius and $\beta_{\rm rec}$ represents the speed at which plasma flows into the reconnection layer. The observations show that the corona of NGC 7469 has an X-ray luminosity of $L_{2-10\,\rm{keV}} = 1.5\times 10^{43}\,\rm{erg\,s^{-1}}$, an X-ray photon index of $s_X = 1.9$, and a cut-off energy of 284 keV \citep{Prince_2025MNRAS.tmp..960P}. 
We can then determine the magnetic field within the corona reconnection layer as
\begin{equation}
\begin{aligned}
    &B = \sqrt{\frac{2L_X}{\beta_{\rm rec} c R_c^2 }}\sim 7.5\times10^3\, \rm{G}\\
    &\left(\frac{L_X}{1.5\times 10^{43} \rm{erg\,s^{-1}}}\right)^{1/2}\left(\frac{0.1}{\beta_{\rm rec}}\right)^{-1/2}\left(\frac{10 R_{\rm g}}{R_c}\right)^{-1}.
  \label{Eq:B}
\end{aligned}
\end{equation}

Our understanding of particle acceleration in relativistic reconnection has been greatly advanced in recent years by first-principles particle-in-cell (PIC) simulations\citep{Werner2017ApJ...843L..27W,Zhang2021ApJ...922..261Z,Zhang2023ApJ...956L..36Z,Chernoglazov2023ApJ...959..122C,Comisso2024ApJ...972....9C,Sironi2015MNRAS.450..183S}. Particle energization is seen to proceed in several stages. At early times, particles gain energy rapidly through direct drift acceleration along the reconnection electric field. Once they reach a critical energy scale,  they are injected into plasmoids and undergo Fermi-like acceleration, where the energy gain per cycle becomes smaller and is regulated by escape and diffusion \citep{Werner2017ApJ...843L..27W,Zhang2021ApJ...922..261Z,Zhang2023ApJ...956L..36Z,Chernoglazov2023ApJ...959..122C,2023ApJ...954L..37L,Comisso2024ApJ...972....9C,Sironi2015MNRAS.450..183S}.


In the sub-relativistic regime ($\sigma \lesssim 1$), {or in the presence of a non-negligible guide field}, reconnection produces particle spectra that are nearly thermal at low energies with a soft non-thermal tail beyond a spectral break, $s\gtrsim 3$. Such a proton spectrum is inconsistent with the hard neutrino spectrum of NGC 7469  (as will be discussed later). On the other hand, protons accelerated by relativistic magnetic reconnection ($\sigma\gtrsim 1$) typically develop a hard borken power-law spectrum, both the pre-break and post-break slopes harden progressively with increasing magnetization $\sigma$, and the break energy also shifts to higher values. {We thus propose that the relativistic magnetic reconnection scenario may apply to NGC 7469 and consider this scenario in the following.}

{In line with simulation results, we assume that a} fraction $f_{\rm nth}$ of the thermal protons in the corona enter the reconnection layer and get accelerated into relativistic energies. The total number density of accelerated protons is thus expressed in terms of the bulk proton density $n_p$ as  
\begin{equation}
    N_p = f_{\rm nth}\,n_p \simeq \frac{f_{\rm nth}B^2}{4\pi\sigma c^2}. 
    \label{Eq:fnth}
\end{equation}
In relativistic reconnection regime, assuming a pre-break spectral index of $s_{\rm pre}\sim1$ and a post-break index of $s_{\rm post}\sim2$, we can write the accelerated proton spectrum, in the absence of cooling, as
\begin{equation}
    \frac{dN_p}{dE_p} \propto \left\{
    \begin{array}{ll}
    E_p^{-1} & m_pc^2< E_p < E_{p,{\rm br}} \\
    E_p^{-2} &  E_{p,{\rm br}}< E_p < E_{p,{\rm max}}
\end{array}
\right.\, .
\label{Eq:proton spectrum}
\end{equation}
We also assume that rough energy equipartition between magnetic fields, X-ray photons, and non-thermal protons is established in the reconnection outflow regions. Correspondingly, we write $U_p = \eta_p U_B$ \citep{Sironi2015MNRAS.450..183S,Petropoulou2019ApJ...880...37P}, where $U_p$ is the energy densities of accelerated protons  and $\eta_p$ is the equipartition parameter. 
Thus
\begin{equation}
    U_p =  \int_{m_p c^2}^{E_{p,{\rm max}}}E_p\frac{dN}{dE_p}dE_p = \eta_p B^2/8\pi.
    \label {Eq:Energy Equipartition}
\end{equation}
Meanwhile, the number density of the non-thermal relativistic protons is given by
\begin{equation}
\begin{aligned}
    N_p &= \int_{m_pc^2}^{E_{p,{\rm max}}}\frac{dN}{dE_p}dE_p \\
    &\approx \frac{\eta_p B^2}{8\pi E_{br}}\approx  10^5\, {\rm cm^{-3}} \left(\frac{B}{7.5\times 10^{3}\,{\rm G}}\right)^2\left(\frac{10\,{\rm TeV}}{E_{br}}\right)\left(\frac{\eta_p}{0.5}\right),
    \label{Eq:N_p}  
\end{aligned}
\end{equation}
where we have substituted the Eq.\eqref{Eq:Energy Equipartition} to replace the normalization. 

Henceforth, we can deduce the break energy of the proton spectrum from Eq.\eqref{Eq:N_p} and Eq.\eqref{Eq:fnth} as 
\begin{equation}
\begin{aligned}
    E_{p,{\rm br}} 
    &\simeq \frac{\eta_p}{2f_{\rm nth}}\sigma m_p c^2 = 2.5  \sigma m_p c^2 \left(\frac{\eta_p}{0.5}\right)\left(\frac{0.1}{f_{\rm nth}}\right).
    \label{Eq:Epbr}
\end{aligned}    
\end{equation}

If the corona of NGC 7469 is composed of an electron-proton plasma, characterized by the condition $n_p = n_e $, the magnetization parameter is $\sigma \approx 0.5 \left({B}/{7.5 \times 10^3 \,{\rm G}}\right)^2\left({R_c}/{10 R_{\rm g}}\right)$, representing a sub-relativistic reconnection regime. In such regime, the proton spectrum exhibits quasi-thermal spectrum with a steep non-thermal tail of $s_{p} \gtrsim 3$ \citep{2023ApJ...954L..37L,Comisso2024ApJ...972....9C,Mbarek2024PhRvD.109j1306M} and a break energy at $\sim 1\, {\rm GeV}$ estimated by Eq.\eqref{Eq:Epbr}. Such a proton spectrum with a peak at GeV is inconsistent with the detection of 100 TeV neutrinos with a rather hard spectrum from NGC 7469. Therefore, an equal proton-electron composition can be ruled out for the corona of NGC 7469. 

Explaining the  neutrino spectrum of NGC 7469  requires either a hard post-break slope ($s_{\rm post} \lesssim 2$) or a high break energy ($E_{p,{\rm br}}\gtrsim 2\,{\rm PeV}$), in which case the neutrino spectrum derives from the hard spectral part of the proton spectrum with $s_{\rm pre} \sim 1$.
In the former case, to achieve a post-break spectral index of $s_{\rm post} \lesssim 2$, the magnetization parameter should satisfy $\sigma \gtrsim 10$ \citep{Mbarek2024PhRvD.109j1306M}.

The allowed proton number density is  $n_p \sim 3\times10^8 \,{\rm cm}^{-3}(B/7.5\times10^3\,{\rm G})^2$ . Then, we can obtain $E_{p,{\rm br}} = 4\,{\rm GeV}\, (0.1/f_{\rm nth})(\eta_p/0.5) $  from Eq.\eqref{Eq:Epbr}. 
In the latter case, adopting $E_{p,{\rm br}}\gtrsim 2\, {\rm PeV}$ to satisfy the 100 TeV neutrino energy {requires a } proton magnetization of $\sigma \gtrsim 10^5$ $(f_{\rm nth}/0.1)(0.5/\eta_p)(E_{p,{\rm br}}/2\,{\rm PeV})$. We deem this value of $\sigma$ is too extreme to be realistic. 

\begin{figure}
    \centering
    \includegraphics[width=0.45\textwidth]{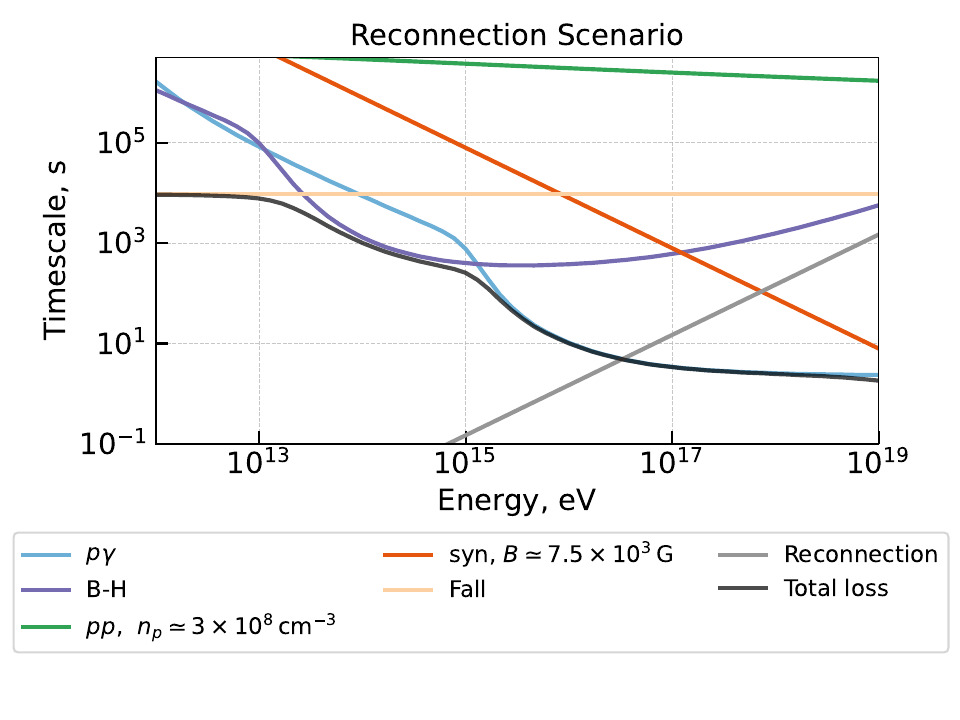}
    \caption{Same as Fig.\ref{Fig:Timescale_Turbulence} but for the reconnection scenario. The parameter values used are $R_c\approx10R_{\rm g}$, $\sigma = 10$, $B = 7.5\times 10^{3}\,{\rm G}$ and $n_p = 3\times 10^8\,{\rm cm^{-3}}$}
    \label{Fig:Timescale_Reconnection}
\end{figure}

\subsection{Maximum proton energy} \label{Sec:Maximum proton energy}

The characteristic magnetic reconnection acceleration timescale \citep{Zhang2021ApJ...922..261Z,Zhang2023ApJ...956L..36Z} is given by
\begin{equation}
    t_{\rm rec} \approx \frac{ r_{\rm L}}{\beta_{\rm rec}c} \approx 4\, {\rm s} \left(\frac{0.1}{\beta_{\rm rec}}\right)\left(\frac{E_p}{10\, \rm{PeV}}\right)\left(\frac{7.5\times 10^3\, {\rm G}}{B}\right),
    \label{Eq:MR Acc}
\end{equation}
where $r_{\rm L} = E_p/eB$ is the Larmor radius, and $\beta_{\rm rec}$ represents the speed at which plasma flows into the reconnection layer, normalized as a fraction of the speed of light.
As we discussed in Appendix.\ref{Sec:Appendix A}, the $p\gamma$ process dominates the cooling within the corona, whereas other cooling processes can be ignored in the high energy band, as shown in Fig.\ref{Fig:Timescale_Reconnection}. 

For this condition, by equating the timescale of reconnection acceleration Eq.\eqref{Eq:MR Acc} with the $p\gamma$ cooling time (Eq.\eqref{Eq:timescale pgamma}), we can derive  the maximum proton energy in the reconnection scenario, 
\begin{equation}
\begin{aligned}
    E_{p,{\rm rec,max}} &=\frac{\beta_{\rm rec} e B}{n_{\gamma}\sigma_{p\gamma}\kappa_{p\gamma}} \simeq  60\,{\rm PeV}   \left(\frac{\beta_{\rm rec}}{0.1}\right)\left(\frac{B}{7.5\times 10^3 G}\right) \\
    &\left(\frac{2\times 10^{39} \,{\rm erg\,s^{-1}}}{L_{\varepsilon_{\gamma}}}\right)\left(\frac{\varepsilon_{\gamma}}{2.5\,{\rm eV}}\right)\left(\frac{R}{10R_{\rm g}}\right),
    \label{Eq:Ep rec max}
\end{aligned}
\end{equation}
where we adopt $\varepsilon_\gamma \sim 0.15\,{\rm GeV}^2/E_{p,{\rm rec,max}} \simeq 2.5 \,{\rm eV}$ as the characteristic energy of the target photons interacting with protons at the maximum energy.  The corresponding photon field luminosity at $\varepsilon_\gamma \simeq 2.5 \,{\rm eV}$ is $L_{\varepsilon_{\gamma}} \simeq 2\times 10^{39}\,{\rm erg\,s^{-1}}$ , which is adopted from the soft photon field of “Big Blue Bump” component as characterized by
\cite{Prince_2025MNRAS.tmp..960P}.
Accordingly, the maximum neutrino energy, $E_{\nu,{\rm rec, max}}\simeq E_{p,{\rm rec, max}}/20 \simeq 2\,{\rm PeV} $, is evidently sufficient to explain the observed neutrino alert associated with NGC 7469.

\subsection{Neutrino spectrum}
We now investigate whether the neutrino signals observed by IceCube in coincidence with NGC 7469 can be explained by the reconnection acceleration, focusing on neutrinos produced via $p\gamma$ process. 
In our parameter space of magnetization, we examine the case of $\sigma \simeq 10$ with the corresponding proton number density of $n_{p} \simeq 3 \times 10^8\,{\rm cm^{-3}}(B/7.5\times10^3\,{\rm G})^2$, ensuring the post-break proton spectral index with $s\simeq2$. Similarly as in the turbulence scenario, we also consider two cases with a high neutrino flux  of $10^{-12}\, {\rm erg\,cm^{-2}\,s^{-1}}$ and a low neutrino flux  of $10^{-13}\, {\rm erg\,cm^{-2}\,s^{-1}}$.
These two cases with different level of neutrino flux may imply different values for the equipartition parameter $\eta_p$ of the corona. Namely, we take $\eta_p \simeq 0.2$ for the high neutrino flux and $\eta_p \simeq 0.02$ for the low neutrino flux, respectively. The variation of $\eta_p$ changes the proton energy density $U_p$, but it is compensated by the modification of the break energy, so that the non-thermal proton number density $N_p$ remains constant.

\begin{figure}
    \centering
    \subfigure{
    \includegraphics[width=0.45\textwidth]{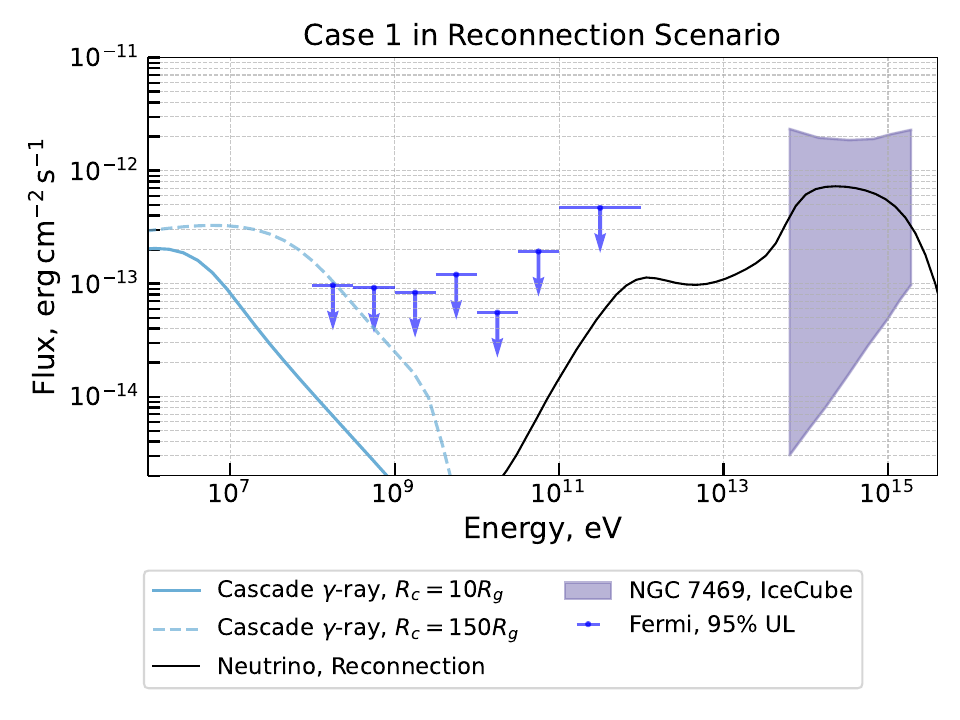}}
    \subfigure{
    \includegraphics[width=0.45\textwidth]{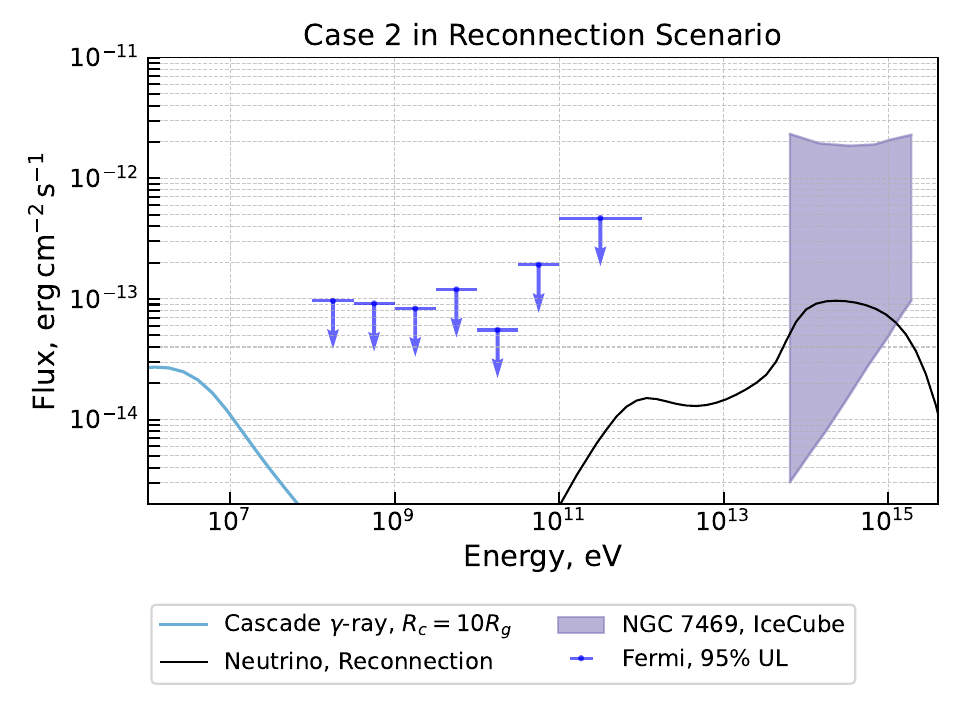}}
\caption{Same as Fig.\ref{Fig:Neutrino spectrum turbulence} but for the reconnection scenario. The upper panel illustrated the case 1 with the high neutrino flux of $10^{-12}\, {\rm erg \,cm^{-2}\,s^{-1}}$ corresponding to $\eta_p\simeq 0.2$, while the lower panel shows the case 2 with a low neutrino flux of $10^{-13} \, {\rm erg \,cm^{-2}\,s^{-1}}$, corresponding to $\eta_p\simeq 0.02$.  
The parameter values used are $\sigma = 10$, $n_p\simeq 3\times 10^8\, {\rm cm^{-3}}$, and $ E_{p,{\rm max}}\simeq60 \,{\rm PeV}$.}
    \label{Fig:Neutrino spectrum reconnection}
\end{figure}

The two panels in Fig.\ref{Fig:Neutrino spectrum reconnection} present the neutrino spectrum  and the related cascade gamma-ray spectrum in two cases.  In both cases, the neutrino spectrum with the proton magnetization $\sigma \sim 10$  is shown in black solid line. The break energy in case 1 and case 2, estimated from Eq.~\eqref{Eq:Epbr}, are $E_{p,{\rm br,1}}\sim 1.5\,{\rm GeV}(0.1/f_{\rm nth}) (\eta_p/0.2)$ and $E_{p,{\rm br},2}\sim 0.15\,{\rm GeV}(0.1/f_{\rm nth}) (\eta_p/0.02)$, respectively.  In addition, similar to the turbulence scenario, the upper-limit GeV flux imposed by {\it Fermi}-LAT constrains the size of the neutrino-emitting region to be $R_c < 150R_{\rm g}$ in case 1 by ensuring that the cascade emission (shown by the blue dashed line) associated with neutrino emission does not exceed the observed upper limit.

\section{Conclusions and Discussions}
In this work, we first analyzed the {\it Fermi}-LAT data of NGC 7469, which yields non-detection. The upper limit of the GeV flux can constrain the size of the emitting region of neutrinos to be less than $150R_{\rm g}$ for a neutrino flux at the level of $10^{-12}\, {\rm erg \,cm^{-2}\,s^{-1}}$. Then, we investigated whether the high-energy neutrinos  from NGC 7469 can originate from its corona based on the turbulence or reconnection acceleration of relativistic protons. Two 100 TeV neutrinos from NGC 7469 in the absence of TeV neutrinos indicate a hard neutrino spectrum, which suggests a hard proton spectrum as well as a large maximum proton energy of $E_{p,{\rm max}}\gtrsim 2 \,{\rm PeV}$.  

In the turbulence acceleration scenario, in order to account for neutrinos with energies up to hundreds of TeV, the corona should be highly magnetized, with a magnetization $\sigma \sim 1$, corresponding to $\beta_p\sim 0.1$. {Particle acceleration becomes so efficient than backreaction of accelerated particles on the turbulence becomes unavoidable on long timescales. The energy distribution takes a broken powerlaw shape with index $s\simeq 2$ at the highest energies. However, the comparatively low neutrino luminosity of NGC~7469 implies that the parent protons have a number density well below that of the electrons. These results and constraints can be accommodated if the corona composition is dominated by pairs, with $n_p/n_e \sim O(10^{-2})$, or if the region where particle acceleration occurs to PeV energies has a comparatively small filling fraction in the coronal volume.}

In the reconnection acceleration scenario, 
a strongly magnetized corona enables efficient particle acceleration, yielding a hard proton spectrum that can be described by a broken power law with a pre-break index of $s_{\rm pre} \sim 1$ and a post-break index of $s_{\rm post} \gtrsim 2$. Both the break energy and the spectral slopes are tightly linked to the magnetization parameter $\sigma$. To produce a hard neutrino spectrum with $s_{\rm post} \simeq 2$,  $\sigma \sim 10$ is required, corresponding to a proton number density of $n_p \lesssim 3\times10^8 \,{\rm cm^{-3}}(B/7.5\times10^3\,{\rm G})^2$. {Here as well,} this value is significantly lower than the pair density, $n_e \simeq 10^{11}\,{\rm cm^{-3}}(10R_{\rm g}/R_c)$, indicating that the corona is predominantly pair-dominated.

By contrast, the case of NGC 1068 differs, as the observed neutrino signal, with a lower peak energy and a soft spectral index ($s_\nu \sim 3.4$), suggests  an intrinsically soft CR spectrum with $s \sim 3$, with a tight energy budget requiring that the high-energy proton luminosity exceeds several percents of the Eddington luminosity. These features can be accommodated
in a proton-electron corona ($n_p\sim n_e$) with moderate magnetization $\sigma \sim 0.1$\citep{Murase2020PhRvL.125a1101M,Inoue2019ApJ...880...40I,Fiorillo2024ApJ...974...75F,Eichmann2022ApJ...939...43E,2025A&A...697A.124L}. 
Our results thus indicate that neutrino observations provide unique probes of the particle acceleration mechanism and of the physics of AGN corona. 

\section{Acknowledgment}
We would like to thank Shiqi Yu and Elisa Resconi for helpful discussions. This work
is supported by the National Natural Science Foundation of
China (grant Nos. 12333006 and 12121003, 12393852). We
are grateful to the High Performance Computing Center of
Nanjing University for doing the numerical calculations in this
paper on its blade cluster system. The work of M.L. is supported by the French Agence Nationale de la Recherche, ANR, project ANR-25-CE31-3279. 

{\em Note added.} While we were finalizing this manuscript, we
became aware of the work of \cite{Salvatore2025arXiv250917751S} (2025, arXiv:2509.17751), which also
analyzed the {\it Fermi}-LAT data of NGC 7469.

\appendix
\section{Cooling and escape processes of protons in the corona} \label{Sec:Appendix A}

For the cooling processes of cosmic ray protons, we consider inelastic collisions ($pp$), photomeson production ($p\gamma$), Bethe-Heitler pair production (B-H), and proton synchrotron radiation. For escape terms, we consider diffusion and free-fall (infall to the BH) as the primary escape processes. 

Firstly, we consider the $p\gamma$ process and Bethe–Heitler pair production. The soft photon field serving as the scattering target can be divided into two components. The first component is the optical–ultraviolet (OUV) photon field, which manifests as the so-called Big Blue Bump (BBB) in AGN spectra and originates from the accretion disk. The "BBB" emission of NGC 7469 has been discussed in detail by \cite{Prince_2025MNRAS.tmp..960P}.   Here, the OUV photon field is modeled as the multi-temperature blackbody emission of the entire accretion disk;  The second component is the X-ray emission produced by the magnetized corona. The observations show that the corona of NGC 7469 has an X-ray luminosity of $L_{2-10\,\rm{keV}} = 1.5\times 10^{43}\,\rm{erg\,s^{-1}}$, an X-ray photon index of $s_X = 1.9$, and a cut-off energy of 284 keV.  Thus the timescale of $p\gamma$ is shown by Eq.\eqref{Eq:timescale pgamma}, calculated by
\begin{equation}
     t_{p\gamma} \approx  110\, {\rm s}\left(\frac{10^{41} \,{\rm erg\,s^{-1}}}{L_{\varepsilon_{\gamma}}}\right)\left(\frac{\varepsilon_{\gamma}}{75\,{\rm eV}}\right)\left(\frac{R}{10R_{\rm g}}\right).
    \label{Eq:timescale pgamma calculated} 
\end{equation}

and the Bethe-Heitler timescale $t_{\rm B-H}$  is
\begin{equation}
\begin{aligned}
    t_{\rm B-H}\approx (n_{\gamma}\hat{\sigma}_{\rm B-H}c)^{-1} \simeq 360\,{\rm s}\left(\frac{10^{41} \,{\rm erg\,s^{-1}}}{L_{\varepsilon_{\gamma}}}\right)\left(\frac{\varepsilon_{\gamma}}{75\,{\rm eV}}\right)\left(\frac{R}{10R_{\rm g}}\right),
    \label{Eq:timescale BH}
\end{aligned}    
\end{equation} 
where $\hat{\sigma}_{\rm B-H}\approx 0.8\times10^{-30} \rm{cm}^{2}$ is the effective cross section for the Bethe-Heitler process. $n_{\gamma} = L_{\varepsilon_{\gamma}}/4\pi R^2c\varepsilon_{\gamma} $ is the number density of the target photon field with $L_{\varepsilon_\gamma}$ the luminosity at photon energy $\varepsilon_\gamma$.  Here $\varepsilon_\gamma = 75\, {\rm eV}$ corresponds to the proton energy of $E_p = 2\, {\rm PeV}$, calculated by the $E_p\varepsilon_{\gamma} \simeq 0.15 \, {\rm GeV}^2$.

Then we try to determine the timescale of $pp$ process. The timescale of $pp$ collision is shown by

\begin{equation}
\begin{aligned}
    t_{pp} \approx (n_{p}\hat{\sigma}_{pp}\kappa_{pp}c)^{-1} \sim 10^5\,{\rm s}
    \left(\frac{10^{10}\,{\rm cm}^{-3}}{n_p}\right),
\label{Eq:timescale pp}
\end{aligned}
\end{equation}
where $\hat{\sigma}_{pp} \simeq 4\times 10^{-26} \rm{cm}^{2}$ and $\kappa_{pp} \approx 0.5$ are cross section and inelasticity for $pp$ process, respectively \citep{Kelner2006PhRvD..74c4018K}.  The $pp$ process is found to be sub-dominant for both cooling and neutrino production in both turbulence and reconnection scenarios, and can thus be safely neglected.

The proton synchrotron timescale is
\begin{equation}
    t_{p,\rm{syn}} = \frac{6 \pi m_p c}{\gamma_p \sigma_T B^2} \left( \frac{m_p}{m_e} \right)^2
    \approx 10^3 \,{\rm s}\left(\frac{60\,{\rm PeV}}{E_p}\right)\left(\frac{7.5\times 10^3\,{\rm G}}{B}\right)^2.
\label{Eq:timescale_syn}
\end{equation}

The timescales of free-fall is $t_{\rm fall} \approx R/V_{\rm R}$. $V_{\rm R} \simeq \alpha V_k/2$ is the radial velocity where $V_k$ is Keplerian velocity and $\alpha\simeq 0.3$ is viscous parameter. In the turbulence scenario,
Particle transport occurs via scattering on magnetic inhomogeneities (mean free path $\lambda_{\rm scatt}$) and turbulent diffusion ($\kappa_{\rm turb} \sim cv_A/3$). Since $\lambda_{\rm scatt} \sim r_L^{1/3}\ell_c^{2/3}$, escape is mainly governed by turbulence. We therefore adopt $\kappa = \kappa_{\rm turb}  + \lambda_{\rm scatt}$c/3,  giving an diffusion escape timescale $t_{\rm diff} = R_c^2/(2\kappa)$.

The total cooling timescale is $t_{\rm cool}^{-1} = t_{p,{\rm syn}}^{-1}+t_{pp}^{-1}+t_{\rm B-H}^{-1}+t_{p\gamma}^{-1}$  and the total escape timescale is calculated as  $  t_{\rm esc}^{-1} = t_{\rm diff}^{-1}+ t_{\rm fall}^{-1}$. Hence, we can calculate the timescale of the total energy loss as
\begin{equation}
    t_{\rm loss}^{-1} = t_{\rm cool}^{-1}+ t_{\rm esc}^{-1}.
\label{Eq:loss}
\end{equation}

Therefore, by utilizing the critical  $p\gamma$ and Bethe-Heitler expressions and all the other timescale formulas above, we can plot the timescales in the corona of NGC 7469 as shown in Fig.\ref{Fig:Timescale_Turbulence} and Fig.\ref{Fig:Timescale_Reconnection} for reconnection regime and turbulence regime, respectively. For protons with energies below 10 TeV, energy loss is primarily governed by escape. For protons with high energy levels below 500 TeV, the Bethe-Heitler (B-H) process serves as the primary cooling mechanism. When proton energies exceed the PeV threshold, the $p\gamma$ process becomes the dominant mechanism, facilitating both proton cooling and neutrino production. The maximum energy of protons, which is controlled by the balance of $p\gamma$ cooling and acceleration. Without efficient $pp$ interactions, the neutrino spectrum lacks a flat low-energy component, which may explain the absence of 1–10 TeV neutrinos from the direction of NGC 7469 as reported by IceCube. Consequently, the $p\gamma$ channel becomes the dominant neutrino production mechanism, but it is only efficient once the proton energy exceeds the PeV scale.

\bibliography{main.bib}{}
\bibliographystyle{aasjournal}
\end{document}